# Thermal Resistances of Thin Films of Small Molecule Organic Semiconductors

Y. Yao[a], Maryam Shahi[a], Marcia M. Payne[b], J.E. Anthony[b], and J.W. Brill[a]

Departments of (a) Physics and Astronomy and (b) Chemistry
University of Kentucky, Lexington, KY 40506-0055

**Abstract:** We have measured the thermal resistances of thin films of the small molecule organic semiconductors bis(triisopropylsilylethynyl) pentacene (TIPS-pn), bis(triethylsilylethynyl) anthradithiophene (TES-ADT) and difluoro bis(triethylsilylethynyl) anthradithiophene (diF-TES-ADT). For each material, several films of different thicknesses have been measured to separate the effects of intrinsic thermal conductivity from interface thermal resistance. For sublimed films of TIPS-pn and diF-TES-ADT, with thicknesses ranging from < 100 nm to > 4 μm, the thermal conductivities are similar to that of polymers and over an order of magnitude smaller than that of single crystals, presumably reflecting the large reduction in phonon mean-free path in the films. For thin (≤ 205 nm) crystalline films of TES-ADT, prepared by vapor-annealing spin-cast films, the thermal resistances are dominated by interface scattering.

## Introduction

In recent work, the thermal conductivities (κ) of layered crystals of several small molecule organic semiconductors have been reported.[1-6] For molecules with planar backbones and silylethynyl (or germanylethynyl) sidegroups projecting between planes, very high interplanar thermal conductivities have been observed.[4,5] For example, while the in-plane, "needle-axis" (i.e. the direction of crystal growth and highest electronic conductivity) thermal conductivity of bis(triisopropylsilylethynyl) pentacene (TIPS-pn[7-10]) $\kappa_{needle} \approx 1.6$ W/m·K,[4] the inter-plane (c-axis) thermal conductivity has been measured, using an ac-photothermal technique,[5] to be $\kappa_c \approx 21$ W/m·K, a value close to that of sapphire.[11] As discussed below, this large value and inverted anisotropy (i.e. $\kappa_c > \kappa_{needle}$) has tentatively been associated with heat flowing between layers via interactions between librations of alkyl chains terminating the silylethynyl sidegroups on the molecules.[5] In contrast, rubrene, with tetracene backbones and much more rigid phenyl sidegroups, has an interlayer thermal conductivity of only $\kappa_c \approx 0.07$ W/m·K $\approx \kappa_{needle}/6$.[3]

While such large thermal conductivities would provide efficient dissipation of Joule heat and therefore bode well for electronic applications of these materials, most organic semiconducting devices require materials in thin film rather than bulk crystal form. Thin film thermal resistances, even for crystalline films, can be much larger than the values deduced from bulk crystalline conductivities, either because of reduced mean-free paths in the material due to increased disorder or because of interfacial thermal resistance with the substrate. Therefore, it was desirable to measure the thin film thermal resistance of TIPS-pn and other small molecule materials being considered for electronic applications.

In this paper, we report on the thin film thermal resistances of films of TIPS-pn and two materials with similar sidegroups and crystal structures, bis(triethylsilylethynyl) anthradithiophene (TES-ADT[7,12,13]) and difluoro bis(triethylsilylethynyl) anthradithiophene (diF-TES-ADT[14,15]), on sapphire and thermally oxidized silicon substrates. (Because of small and irregularly shaped crystals, bulk crystal values of the thermal conductivities of TES-ADT and diF-TES-ADT have not yet been reported, but our preliminary photothermal measurements on a TES-ADT crystal with a non-uniform thickness gave a value $\kappa_c \approx (5 \pm 2)$ W/m·K, a few times smaller than the value for TIPS-pn but still very large.) For these thin film measurements, we use the well-established 3ω-technique,[1,16-22] which, as described below, yields values of the film thermal resistance

$$R_{film} = (t/\kappa + \rho_{int})/S, \quad (1)$$

where t and S are the thickness and area of the film, $\rho_{int}$ is the interfacial thermal resistivity, and κ is the through-plane thermal conductivity (i.e. for layered, crystalline films, $\kappa = \kappa_c$). Therefore, to separate the interfacial from intrinsic resistivity, it is useful to measure the thickness dependence of $R_{film}$ which, as emphasized in Ref. 6, has not always been done for organic semiconductor films.[1,17-20] For example, for a 100 nm thick film of a material with $\kappa \approx 0.3$ W/m·K (similar to many polymers[19-21]), the two terms in Eqtn. (1) are comparable for $\rho_{int} \approx 3 \times 10^{-7}$ m$^2$K/W. This value is only about an order of magnitude larger than that of evaporated metal,[23,24] oxide,[25] or organometallic[26] films on ceramic or metal substrates, close to the value of evaporated polycrystalline pentacene films on silicon oxide,[6] and an order of magnitude smaller than the interface resistances of the best epoxy resins[27] or thermal greases.[28]

The samples measured are "vapor annealed"[12,13,29] and non-annealed spin-cast films of TES-ADT, with thicknesses ranging from 77 nm to 205 nm, and sublimed films of TIPS-pn and diF-TES-ADT, with thicknesses ranging from 100 nm to 4 μm. As described below, the vapor annealed films are crystalline, with mm sized crystallites oriented with **c** in the through-plane direction,[12,13,29] while the other films are thought to be *ab*-plane disordered, but with the molecular sidegroups also largely oriented through-plane.[10,29,30] We find that the thermal resistances are much larger than expected from the high single crystal thermal conductivities, and suggest that the reason is due to disorder and significant interface thermal resistances in the films.

## Experimental
### Technique

The 3ω−technique for thermal conductivity measurements uses a single metal strip (length = L and width = w) deposited on the sample surface to act as both a resistive heater and thermometer.[31] An ac driving current (at frequency ω) is applied to the heater, so that its temperature (and the temperature of the nearby substrate) will oscillate at frequency 2ω. Consequently, its electrical resistance will oscillate at 2ω, producing a third harmonic ($V_{3\omega}$) in the voltage drop across the metal strip, with $V_{3\omega} \propto \Delta T$, the magnitude of its (in-phase) temperature oscillation. (See Eqtn. (3) below.) We use a sensitive bridge circuit and a lock-in amplifier (with differential input) to remove the drive voltage (including its third harmonic distortion), so that the 3ω signal can be used to infer the magnitude of the temperature oscillations and therefore the thermal response of the substrate:[16,31]

$$\Delta T = P/(2L\pi\kappa_{sub}) [-\ln(\omega)+\ln(\kappa_{sub}/c_{sub}w^2)+1.74]. \quad (2)$$

Here P is the applied power and $c_{sub}$ and $\kappa_{sub}$ are the specific heat per unit volume and thermal conductivity of the substrate. ΔT was determined from $V_{3\omega}$ using[31]

$$\Delta T = 2V_{3\omega}/\alpha V_0, \quad (3)$$

where $V_0$ was the voltage across the heater (measured in a 4-probe configuration) and $\alpha \equiv (1/R)\,dR/dT$ the temperature coefficient of resistance for the copper strip. We found $\alpha$ by measuring the dc resistance of the heater as its temperature was slowly heated and cooled (~ 30 $^o$C) in the same vacuum cryostat in which the thermal conductivity was measured.

Lee and Cahill[16] have shown that the 3ω−technique could be used to measure the transverse (i.e. through-plane) thermal conductivity of a thin film of thickness t, deposited between the substrate and heater, once the substrate thermal conductivity was known; the (transverse) film thermal resistance just added a frequency independent offset to the $\ln(\omega)$ dependence of the substrate:

$$\Delta T(\text{film}) \equiv \Delta T(\text{film+substrate}) - \Delta T(\text{substrate})$$
$$= Pt/(wL\kappa_{app}) = P/(wL)\,(t/\kappa + \rho_{int}) = PR_{film} \quad (4)$$

where the effective area in Eqtn. (1), $S = wL$, and $\kappa_{app}$ is the apparent thermal conductivity of the film. The basic assumption is that the heat flow in the film is one-dimensional, i.e. w >> t, while, as for bulk measurements, w is much smaller than the thermal diffusion length in the substrate.[16,31] In addition, one needs t << the thermal wavelength in the film[16] and $\kappa_{app} < \kappa_{sub}$.[22]

Copper strips (~ 50 nm thick, L = 6mm long, w = 50μm wide, resistances ~ few hundred ohms) as heaters/thermometers were evaporated through a shadow mask on the samples (bare substrates or thin films). The metal strip requires a fairly large temperature coefficient of resistance to generate a measurable resistance change as a function of temperature. We chose copper for its large temperature coefficient of resistance, ease of evaporation, and low cost. (While the resistance of these films changed slowly and slightly due to oxidation, we corrected for these changes by measuring the temperature dependence of the resistance before and after thermal conductivity measurements.)

**Sample Preparation**

Before deposition of thin films on our substrates (sapphire and thermally oxidized silicon), the substrates were first immersed in an acetone bath and sonicated at room temperature for 5 minutes. To remove the residue of acetone, a similar treatment was applied to the substrate with isopropanol or mixed alcohols followed by deionised water. Finally, the substrate surface was blown with compressed nitrogen until dry. In some cases, the surface was then treated with UV-ozone as a final step, but we found that the latter had insignificant effect on the interface thermal resistances for our samples.

TES-ADT crystallizes very slowly from solution and non-crystalline films with uniform thicknesses form from spin-cast solutions.[12,13,29] To prepare these, TES-ADT was dissolved in toluene to form a 2 wt% solution, which was spin-coated on to the substrate at 1000 rpm. Then the sample was heated in air at 80 $^o$C to remove residual solvent, yielding a ~ uniform thickness film. As described in Ref. 29, molecules in these films are oriented with their *c*-axes normal to the film but are mostly disoriented in the *ab*-plane, although small monoclinic crystallites may be present. Exposing these films to dichloroethane (DCE) vapors for 1 – 10 minutes promotes spherulite growth into triclinic crystallites, typically ~ 1mm in size,[12,13,29] while the thickness stays uniform. The films made by this process are typically 100 nm thick, as measured with an AFM (after the thermal resistance measurement). Unfortunately, thicker uniform films cannot be grown from solution and TES-ADT degrades when evaporated.

On the other hand, while diF-TES-ADT and TIPS-pn precipitate quickly from solution to form non-uniform, granular films,[8,9,30] uniform and thicker films[10] can be prepared by vacuum evaporation,[10] although the uv-visible absorption spectra of the TIPS-pn films indicates that there may be small amounts of degradation products present. The thicknesses of the sublimed films (100 nm to 4 μm), were determined approximately during sublimation by a quartz crystal thickness monitor and more accurately (± 10 nm) with a Dektak 6M profilometer after the thermal measurement. (The two thickness measurements typically agreed within 10%.) We assume that the molecules in the sublimed films have their sidegroups mostly aligned transversely, although the films are presumably microcrystalline and/or disordered in the *ab*-plane, as discussed for evaporated TIPS-pn films in Ref. 10. Similarly, spin-cast diF-TES-ADT films on thermally oxidized silicon were observed to be microcrystalline but with the c-axes ~ 80% aligned perpendicular to the substrate.[30] Supporting our assumption of *ab*-plane disorder is that no grains were visibly present in these films and infrared absorption spectra for small areas [~ (50 μm)$^2$] of films evaporated on (infrared transparent) KRS-5 substrates measured with an infrared microscope showed no polarization dependence. (In fact, the diF-TES-ADT spectra were identical to that of Figure 4a in Reference 15 for diF-TES-ADT dispersed in KBr.)

Two checks were made to insure that evaporation of the copper heater/thermometer film did not damage or otherwise affect the organic film. 1) We exposed organic films to heated tungsten evaporation sources for prolonged periods and checked their thicknesses with the AFM before and after; thicknesses changed less than 10 nm, indicating that exposure to the hot source did not cause significant evaporation of the organic film. 2) Polarized infrared transmission spectra (Figure 1) were taken on vapor annealed TES-ADT films which were deposited on KRS5 substrates, as functions of distance from the copper heater using an IR microscope. The spectra were taken in 100 μm square areas adjacent to the copper line and 1 mm and 2 mm away, but all on the same spherulite. Absorption lines at ~728, 857 and 878 cm$^{-1}$ are strongly polarization dependent. The spectra were independent of distance, indicating that evaporation through the shadow mask did not cause degradation or melting (and realignment) of the TES-ADT spherulites.

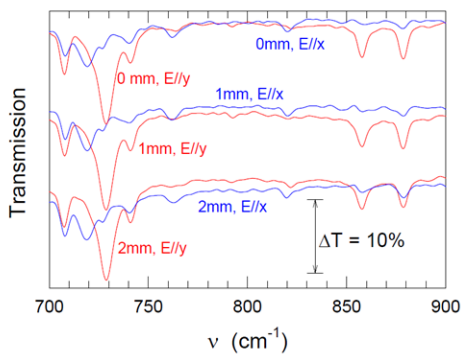

**Figure 1.** Polarized infrared transmission spectra of a single spherulite in a spin-cast, vapor annealed TES-ADT film on a KRS5 substrate, taken in (100 μm)$^2$ spots adjacent to the copper heater line and 1 mm and 2 mm away from the line. (x and y refer to perpendicular microscope axes.). The curves for each position are vertically offset.

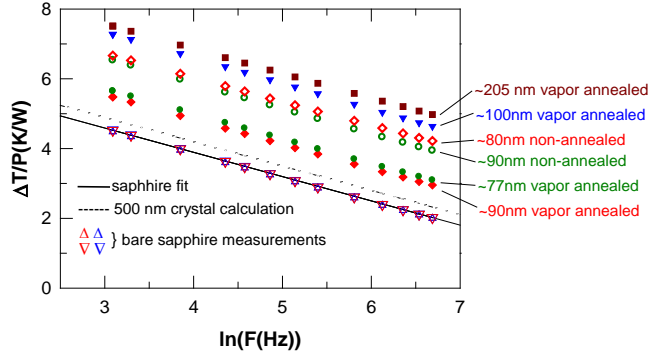

**Figure 2.** Measured frequency dependence of the ΔT/P for spin-cast TES-ADT films on sapphire substrates. The open triangles show the results on bare sapphire and the solid line is a fit to Eqtn. (2). The dashed line shows the expected results for 500 nm thick TES-ADT, assuming $\kappa = \kappa_c$(crystal) = 5W/m·K. The solid symbols show results for vapor annealed films of different thicknesses, as shown, and the open circles and squares show results for non-annealed films.

## Results and Discussion

Figure 2 shows measured values of ΔT/P as a function of frequency ($F = \omega/2\pi$) for several spin-cast TES-ADT films. Also shown (open triangles) are the results for heaters on four bare sapphire substrates; the solid line shows ΔT/P calculated with $\kappa_{sapphire} = 38.9 \pm 0.8$ W/m·K, consistent with the published value.[11] The dashed line shows the calculated value of ΔT/P for a 500 nm thick film of TES-ADT assuming our measured bulk crystal thermal conductivity value $\kappa_c = 5.5$ W/m·K. The solid symbols show our measurements on vapor annealed TES-ADT films made from different starting solutions while the open circles and open diamonds show results for two spin-cast but non-annealed (i.e. non-crystalline) films. The thermal resistance values vary from 1.0 to 2.8 K/W, much larger than expected from the bulk crystal measurements (e.g. as shown by the dashed line), and for the crystalline films doesn't scale with film thickness, suggesting that for these, the thermal resistance is dominated by the interface thermal resistivity; the resulting values of $\rho_{int}$ vary from (3 to 8) x $10^{-7}$ m$^2$K/W, which, as mentioned above, are reasonable values for deposited films. (We will suggest below that the non-annealed films may have lower $\rho_{int}$'s than the vapor annealed films.)

Measurements on spin-cast TES-ADT films (~ 100 nm thick) were also carried out on doped silicon substrates (with thermally oxidized surfaces), the most common substrate for organic thin film transistors. The calculated baseline for silicon, with $\kappa_{Si} = 142$ W/m·K,[32] as well as the measured value on the thermally oxidized substrate are shown in Figure 3, with the measured values for vapor-annealed and non-annealed films prepared at the same time. The thermal resistances are comparable to those measured on sapphire. (The nonlinearities for F > 30 Hz are caused by capacitive coupling of the heater to the conducting, grounded doped silicon.)

The results of 3ω measurements on several sublimed diF-TES-ADT films and TIPS-pn films of different thicknesses on sapphire are shown in Figure 4. Note that for the thicker films, t approaches $(D/2\omega)^{1/2}$, the thermal wave length, explaining the downward curvature at high frequencies.

The thickness dependence of the thermal resistances for these sublimed films is shown in Figure 5. Although there is some scatter, presumably reflecting the quality of the films, both materials exhibit a rough linear dependence of $R_{film} \equiv \Delta T_{film}/P$ on thickness. As shown in the inset, the intercepts are very small: $R_{film}(t=0) < 0.6$ K/W corresponding to $\rho_{int} < 2 \times 10^{-7}$ m$^2$K/W. That is, the sublimed films have lower interface resistivities than the annealed, spin-cast TES-ADT films. Since the structures of diF-TES-ADT and TES-ADT are very similar,[13] we expect similar thermal conductivities for the two materials, and in-fact the thermal resistances of the non-annealed spin-cast TES-ADT films, also shown in Figure 5, are consistent with those of the sublimed films of diF-TES-ADT. Therefore, it is possible that like the sublimed films, the non-annealed spin-cast films have low interface thermal resistances, smaller than that of the vapor annealed films. In fact, Lee *et al*[29] have suggested that the vapor annealing process causes dewetting of the films from the substrate, which should increase the interface thermal resistance.

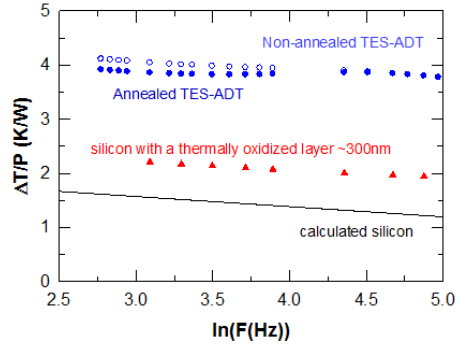

**Figure 3.** Measured frequency dependence of ΔT/P for ≈ 100 nm thick vapor annealed (solid blue circles) and non-annealed (open blue circles) spin-cast TES-ADT films on thermally oxidized silicon. The red triangles show the results on the bare oxidized silicon and the calculated silicon baseline [from Eqtn. (2)] is shown by the solid line.

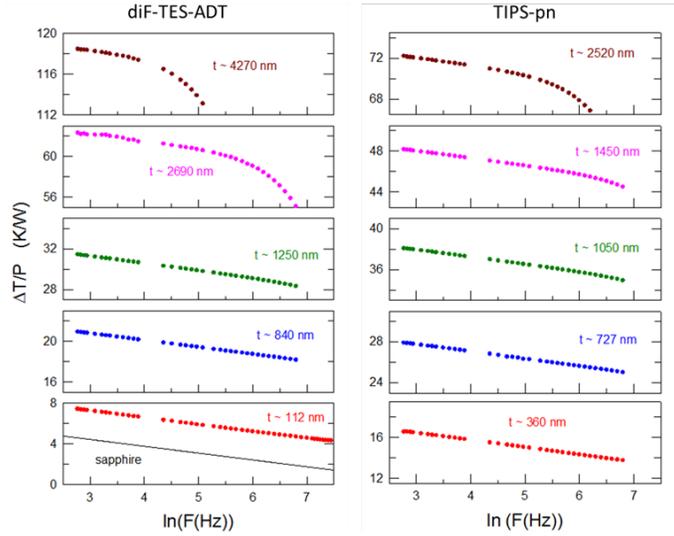

**Figure 4.** Frequency dependence of sublimed films of diF-TES-ADT (left panels) and TIPS-pn (right panels) of the indicated thicknesses on sapphire substrates. The reference bare sapphire line (from Figure 2) is shown in the lower left panel.

For TIPS-pn, the average value of $R_{film}/t$ (using the more accurate profilometer values of thickness) gives $\kappa = (0.104 \pm 0.009)$ W/m·K, where the uncertainty includes the effect of the measurement precision for each film as well as the standard deviation of $R_{film}/t$ values about the average. This value of κ, similar to that of polymers,[18-20] is an order of magnitude smaller than the in-plane (needle axis) thermal conductivity[4] and two orders of magnitude smaller than the *c*-axis[5] thermal conductivity of crystals. Similarly, for diF-TES-ADT, the average value of $R_{film}/t$ gives κ = (0.135 ± 0.015) W/m·K, almost two orders of magnitude smaller than the *c*-axis value for TES-ADT, with a very similar structure.[13]

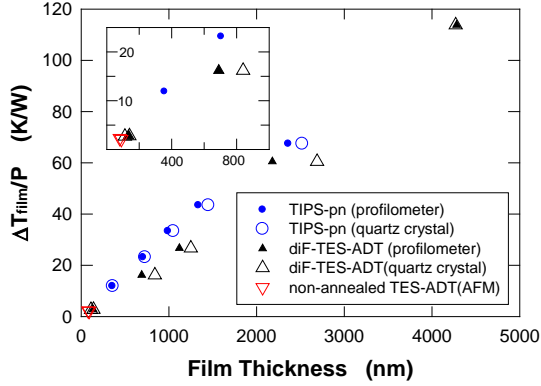

**Figure 5.** Thickness dependence of film thermal resistance ($\Delta T_{film}/P$) for sublimed films of diF-TES-ADT (black triangles) and TIPS-pn (blue circles); the typical uncertainties are ± 0.3 K/W. The solid symbols show the profilometer measurements of the film thicknesses (uncertainties ± 10 nm) while the open symbols show the thicknesses as determined by the quartz crystal monitor during sublimation. Also shown (open inverted red triangles) are the results for the non-annealed spin-cast TES-ADT films. The inset shows a blow-up of the results with t < 1 μm.

In view of the single crystal values, the small values of κ for the sublimed films seem surprising, especially if one assumes that the TIPS and TES sidegroups are still mostly aligned transverse to the substrates. For the single crystals, the large values of $\kappa_c$ were thought to be inconsistent with acoustic phonons carrying most of the heat (the usual assumption in molecular solids) as they would imply unlikely mean-free-paths of several hundred layers. Instead, it was suggested that interactions between sidegroup librations, i.e. low-energy (< $k_B T$, where $k_B$ = Boltzman's constant) optical phonons, might allow these modes to carry a large fraction of the heat.[4,5] Supporting this possibility have been recent low-energy inelastic x-ray scattering experiments,[33] which have indicated that 11 meV optical phonons have significant energy dispersion along c*, corresponding to phonon velocities ~ 2 km/s, similar to the expected values for acoustic modes.[4] The low values of κ in the sublimed films suggests that film disorder greatly reduces the c-axis mean-free paths of propagating acoustic and/or optical phonons. For example, if one assumes that most of the heat in the crystals is carried by acoustic phonons, with an effective room temperature specific heat of $c_{eff}$ ~ 3 $k_B$/molecule and a typical velocity of v ~ 2 km/s, then from[3-5] κ = 1/3 $c_{eff}$ v λ/Ω, where λ = the average mean-free path and Ω the molecular volume, κ ~ 0.1 W/m·K implies λ ~ 2c. Alternatively, if one assumes optical phonons contribute to heat conduction so that $c_{eff}$ is a few times larger, the average mean-free path will be a few times smaller, e.g. λ ~ c/2. Both cases are qualitatively consistent with models of minimum thermal resistance due to disorder.[34,35]

## Conclusion

The thermal conductivities of sublimed films of TIPS-pn and diF-TES-ADT are much smaller than their crystalline values, presumably because the lack of three-dimensional order in the films severely limits the mean-free path of the conducting phonons, including the librational optical phonons proposed to carry much of the heat. On the other hand, the thermal resistances of thin ($\leq$ 205 nm) crystalline films of TES-ADT, prepared by vapor-annealing of spin-cast films, are dominated by their interface resistances, possibly due to dewetting of the film from the substrate during the annealing process. While not excessive, such thermal resistances might limit the utility of these films in electronic devices. It remains to be determined if solution-cast, high electronic mobility, crystalline films of TIPS-pn and diF-TES-ADT, which are too irregular in thickness for our 3$\omega$-measurements, have comparable interface resistances.

## Acknowledgements

We thank Greg Porter (U. Kentucky) for developing the active bridge circuit, Prof. J. Todd Hastings and John Connell (U. Kentucky) for their assistance in preparing the shadow masks and measuring the thicknesses of films, and Prof. Yueh-Lin Loo and Anna Hailey (Princeton U.) for discussions on vapor-annealing of TES-ADT. This research was supported in part by the United States National Science Foundation, Grant No. DMR-1262261 and the United States Office of Naval Research, Grant No. N00014-11-0328.

## References

1. N. Kim, B. Domerq, S. Yoo, A. Christensen, B. Kippelen, and S. Graham, *Appl. Phys. Lett.,* 2005, **87**, 241908.
2. Y. Okada, M. Uno, Y. Nakazawa, K. Sasai, K. Matsukawa, M. Yoshimura, Y. Kitaoka, Y. Mori, and J. Takeya, *Phys. Rev. B*, 2011, **83**, 113305.
3. H. Zhang and J.W. Brill, *J. Appl. Phys.,* 2013, **114**, 043508.
4. H. Zhang, Y. Yao, M.M. Payne, J.E. Anthony, and J.W. Brill, *Appl. Phys. Lett.*, 2014, **105**, 073302.
5. J.W. Brill, M. Shahi, M.M. Payne, J. Edberg, Y. Yao, X. Crispin, and J.E. Anthony, *J. Appl. Phys.*, 2015, **118**, 235501.
6. J. Epstein, W.-L Ong, C. Bettinger, and J.A. Malen, *ACS Appl.Mater. Interfaces*, 2016, **8**, 19168-19174.
7. M.M. Payne, S.R. Parkin, J.E. Anthony, C.-C. Kuo, and T.N. Jackson, *J. Am. Chem. Soc.* 2005, **127**, 986-4987.
8. C.D. Sheraw, T.N. Jackson, D.L. Eaton, and J.E. Anthony, *Adv. Mater.*, 2003, **15**, 2009-2011.
9. J. Chen, D.C. Martin, and J.E. Anthony, *J. Mater. Res.,* 2007, **22**, 701-1709.
10. S.C.B. Mannsfeld, M.L. Tang, and Z. Bao, *Adv. Mater.* 2011, **23**, 127-131.
11. V. Pischik, L.A. Lytvynov, and E.R. Dobrovinskaya, *Sapphire:Material, Manufacturing, Applications*, (Springer, 2009, New York).
12. K.C. Dickey, J.E. Anthony, and Y.-L. Loo, *Adv. Mater.* 2006, **18**, 1721-1726.
13. S.S. Lee, C.S. Kim, E.D. Gomez, B. Parushothaman, M.F. Toney, C. Wang, A. Hexemer, J.E. Anthony, and Y.-L. Loo, *Adv. Mater.,* 2009, **21**, 3605-3609.
14. O.D. Jurchescu, S. Subramanian, R.J. Kline, S.D. Hudson, J.E. Anthony, T.N. Jackson, and D.J. Gundlach, *Chem. Mater.* 2008, **20**, 6733-6737.
15. R.J. Kline, S.D. Hudson, X. Zhang, D.J. Gundlach, A.J. Moad, O.D. Jurchescu, T.N. Jackson, S. Subramanian, J.E. Anthony, M.F. Toney, and L.J. Richter, *Chem. Mater.* 2011, **23**, 1194-1203.
16. S.-M. Lee and D.G. Cahill, *J. Appl. Phys.*, 1997, **81**, 2590-2595.

17. F. Reisdorffer, B. Garnier, N. Horny, C. Renaud, M. Chirtoc, and T.-P Nguyen, *EPJ Web of Conferences*, 2014, **79**, 02001.
18. X. Wang, K.D. Parrish, J.A. Malen, and P.K.L. Chan, *Scientific Reports*, 2015, **5**, 16095.
19. G.-H. Kim, L. Shao, K. Zhang, and K.P. Pipe, *Nature Materials*, 2013, **12**, 719-723.
20. O. Bubnova, Z.U. Khan, A. Malti, S. Braun, M. Fahlman, M. Berggren, and X. Crispin, *Nature Materials,* 2011, **10**, 429-433.
21. T. Borca-Tasciuc, A.R. Kumar, and G. Chen, *Rev. Sci. Inst.*, 2001, **72**, 2139-2147.
22. A. Jacquot, B. Lenoir, A. Dauscher, M. Stolzer, and J. Meusel, *J. Appl. Phys.*, 2002, **91**, 4733-4738.
23. E.T. Swartz and R.O. Pohl, *Appl. Phys. Lett.*, 1987, **51,** 2200-2202.
24. R.J. Stoner and H.J. Maris, *Phys. Rev. B*, 1993, 48, 16373-16387.
25. H.-C. Chieh, D.-J. Yao, M.-J. Huang, and T.-Y. Chang, *Rev. Sci. Inst.*, 2008, **79**, 054902.
26. Y. Jin, Y. Yadav, K. Sun, H. Sun, K.P. Pipe, and M. Shtein, *Appl. Phys. Lett*., 2011, **98**, 093395.
27. K. Fukushima, Y. Takezawa, and T. Adschiri, *Jpn. Jnl. Appl. Phys.*, 2013, **52**, 081601.
28. https://www.dowcorning.com/content/publishedlit/11-1712-01.pdf
29. S.S. Lee, S.B. Tang, D.-M. Smilgies, A.R. Woll, M.A. Loth, J.M. Mativetsky, J.E. Anthony, and Y.-L. Loo, *Adv. Mater*., 2012, **24**, 2692-2698.
30. J.W. Ward, R. Li, A. Obaid, M.M. Payne, D.-M. Smilgies, J.E. Anthony, A. Amassian, and O.D. Jurchescu, *Adv. Funct. Mater*. 2014, **24**, 5052-5058.
31. D.G. Cahill, *Rev. Sci. Inst.*, 1990, **6**1, 802-808.
32. H.R. Shanks, P.D. Maycock, P.H. Sidles, and G. Danielson, *Phys. Rev.*, 1963, **130**, 1743-1748.
33. Y. Yao, J.W. Brill, and A. Alatas, unpublished results.
34. D.G. Cahill, S.K. Watson, and R.O. Pohl, *Phys. Rev. B*, 1992, **46**, 6131-6140.
35. Z. Chen and C. Dames, *Appl. Phys. Lett*., 2015, **107**, 193104.